\documentstyle[epsfig,prb,aps]{revtex}
\begin{document}
\title{
Spin-orbit coupling in the actinide elements: a critical evaluation of 
theoretical equilibrium volumes
}
 
\author{J.M.Wills$^{1}$,  P.H.Andersson$^{2}$, L.Nordstr\"{o}m$^{2}$ P.S\"{o}derlind$%
^{3}$, and O.Eriksson$^{1,2}$}
\address{
1. Theoretical Division,\\
Los Alamos National Laboratory,\\
Los Alamos, New Mexico 87544
}
\address{
2. Department of Physics,\\
Uppsala University\\
Box 530, Uppsala, Sweden
}
\address{
3. Physics Directorate,\\
Livermore National Laboratory,\\
Livermore, California 94550
}
\maketitle
\date{\today}

\begin{abstract}
In a recent paper by Jones {\it et al.}\cite{bs}, it is argued, based on
FP-LAPW band-structure calculations, that
previous calculations of ground-state properties for actinides, using
the FP-LMTO method implemented by J.M. Wills\cite{john}, are in error.
We demonstrate in this paper that the
conclusions of Jones {\it et. al.}\cite{bs} are unfounded. Calculations using the
FP-LMTO method are compared with calculations performed with the FP-LAPW
method and, in contradiction to statements by Jones  {\it et. al.},
they agree very well.
The limitation of the present implementation of the spin-orbit coupling
is
discussed, where especially the 6$p$ states pose a difficulty. The observed
discrepancy (0-10\%) between the FP-LMTO calculations and the recent
FP-LAPW calculation of Jones {\it et al.} is shown to be due to the choice of
muffin-tin radius in the calculations. We argue that the choice of a 
constant muffin-tin
radius, common for all calculated volumes, is less satisfactory compared 
to the choice of a muffin-tin radius that scales with the volume, 
especially when a large volume interval and open packed structures are 
considered.
The conclusion of Jones {\it et al.} that $\alpha $-Pu
has partially delocalized 5$f$ electrons is argued to be erroneous.
\end{abstract}


\section{Introduction}

During the past three decades many first principles calculations of cohesive
properties and electronic structure of the light actinides have been
published
\cite{koelling,brooks,solovyev,picket,vanek,skriver2,ak1,ak2,ak3,ak4,ak5,ak6,ak7,ak8,ak9,ak10,ak11,ak12,ak13,ak14,ak15,ak16,soderlind97,ak18,ak19,ak20,ak21,ak22,ak23,ak24,ak25,ak26,ak27,ak28,ak29}
The calculations have been done with various degrees of sophistication,
ranging from being scalar relativistic and based on the local density
approximation (LDA) within the atomic sphere approximation to fully relativistic, 
full potential calculations based on the generalized gradient approximation (GGA).  
Due to different approximations involved, somewhat different agreement in 
equilibrium volumes and bulk moduli has been obtained, but typically the
calculations underestimate the experimental values of equilibrium volume.
This overestimate of the chemical bond leading to somewhat
small equilibrium volumes occur for all metals in the periodic table
and is a well known shortcoming of the LDA commonly referred to
as the ``LDA contraction''. To some extent the GGA corrects this systematic
error of the LDA and has therefore been a common replacement for the
LDA in recent calculations. Calculations utilizing the GGA for the light
actinide metals results in equilibrium volumes within a few percent of the
experimental data. 
The accuracy of these (LDA and GGA) calculations may best be 
realized by bearing in 
mind that the experimental
volumes change drastically over the series (the volume of the
first actinide, Ac, is roughly twice as large as the last of the light
actinides, $\alpha $-Pu), resulting in drastically different electron densities.
More importantly, an understanding of
the nature of the chemical bonds and electronic structure in the actinides
has emerged
from those calculations, explaining ground-state properties such
as the equilibrium volume, structural
stability, elastic constants, and cohesive energy. The key concepts are 
(i) 5$f$ electrons form band states for the light actinides Th $\to$ $\alpha $-Pu, and
hence they contribute to the chemical bonding according to the 
Friedel model\cite{brooks,skriver}
and (ii) the narrow bandwidth of the 5$f$-electron
states provides a Peierls/Jahn-Teller like symmetry breaking mechanism that
stabilizes open and low symmetry structures in the light actinides.\cite{ak11} 
Based on calculations and
this physical picture it was predicted that  U, Np, and Pu
should undergo phase transitions to more close-packed structures 
such as bcc, hcp, or fcc,\cite{ak11,ak20}
a prediction that has been born out by experiments. This picture outlined above is
also consistent with the experimental observation of a pressure induced bct phases
in Th and Ce-Th alloy systems.\cite{ceth} For these systems theory was in fact able to 
reproduce the experimental data in great detail. In addition, calculations of the structural
behavior of uranium up to 1 Mbar was shown to be in excellent agreement with diamond anvil
cell measurements\cite{diamond}.
This set of results, 
demonstrates that our theoretical
approach is able to produce results that are
comprehensive, detailed, consistent, and, most importantly, explain experimental
low temperature properties of the light actinides.

The present authors have participated\cite{ak3,ak4,ak6,ak7,ak8,ak10,ak11,ak12,ak13,ak14,ak15,ak16,soderlind97,ak18,ak19,ak20,ak21,ak22,ak23,ak24,ak25,ak26,ak27,ak28,ak29} in this theoretical advancement using
a first-principles relativistic, 
full-potential technique implemented using the linear
muffin-tin orbitals method.
A large percentage of these results have been on elemental actinides. In
these studies, the FP-LMTO method has been successfully applied to structural,
cohesive,  magnetic, and elastic properties of actinide elements 
and compounds. These
calculations include both bulk and surface properties. 

In a recent paper, Jones {\it et al.}\cite{bs} argued that the bulk of the 
FP-LMTO
calculations (30 publications or more) mentioned above were "at best
ambiguous" and that there is "an error" in these calculations.
They base this conclusion upon {\it one single fact:}  their calculated equilibrium volumes
for the light actinides are closer, on average the difference is a few percent, 
to experimental data
than the previously published FP-LMTO results. The calculations by Jones {\it et al.} were performed
with two methods, the established FP-LAPW method and 
a, for the heavier elements less used, Gaussian-orbital
method. Some of the calculations were too complex for this latter method and
a complete set of calculations were only presented for the FP-LAPW method in their paper.
For that reason, we will below focus most of our attention on
the FP-LAPW results and make a critical analysis of the possible
sources for the claimed differences between the FP-LMTO and FP-LAPW methods.

\section{Comparing the FP-LAPW and FP-LMTO methods}

The results obtained from FP-LMTO and FP-LAPW calculations have, 
over the years, been compared with
each other and the conclusion has been that these two methods, when carefully used,
produce nearly identical results, independent of the systems that have been studied.
For instance, calculated data for W, published using
FP-LAPW\cite{W-lapw} and FP-LMTO\cite{W-lmto} as well as for published results for
La using FP-LAPW\cite{La-lapw} and FP-LMTO\cite{La-lmto} show that these methods 
give the same results. 
There is absolutely no reason why these two methods should give different results,
they do use different expansions of basis functions but when those are sufficiently converged
the results should be equivalent. 
Therefore, it is most surprising that Jones {\it et al.}\cite{bs} claim that
they
have found an exception to this very well established fact. Without worrying why this
is the case, they instead take the position that all FP-LMTO calculations that have been
performed by the present authors are simply wrong. 
In the present paper we address the
question that Jones {\it et al.}\cite{bs} 
should have done in their paper: why do their FP-LAPW
results
of the equilibrium volumes for the light actinides differ from our published FP-LMTO results?

In examining the results presented by Jones {\it et al.,} it is clear that the element
for which
the FP-LAPW equilibrium volume compares worst with our FP-LMTO equilibrium
volume is Th. The calculations were done employing the same version of the
GGA and 
the spin-orbit coupling was implemented in a similar fashion (only inside the
muffin-tin spheres).
We note that there is a difference in treating the spin-orbit coupling, 
but that the second variation implementation of the spin-orbit interactions should, when carried to completion, give results as accurate as those 
obtained by the more complete FP-LMTO implementation.
The published FP-LMTO result was about 11\% lower than the room temperature equilibrium volume, whereas 
the FP-LAPW calculation by Jones {\it et al.} gave an 
equilibrium volume about 2\% smaller than experiment.
This difference is somewhat surprising and 
motivated us to investigate the reason in detail.
In order to do so, we performed in parallel FP-LMTO and FP-LAPW calculations 
for Th. The
FP-LAPW code is described by Singh\cite{singh} and is actually the same
code that Jones {\it et al.} were using.
Throughout this comparison we performed calculations in as similar fashion as 
possible, concerning exchange/correlation functional, k-space integration, truncations in potential and density 
expansions, choice of muffin-tin radii etc.

\begin{figure}
\begin{center}
\epsfig{file=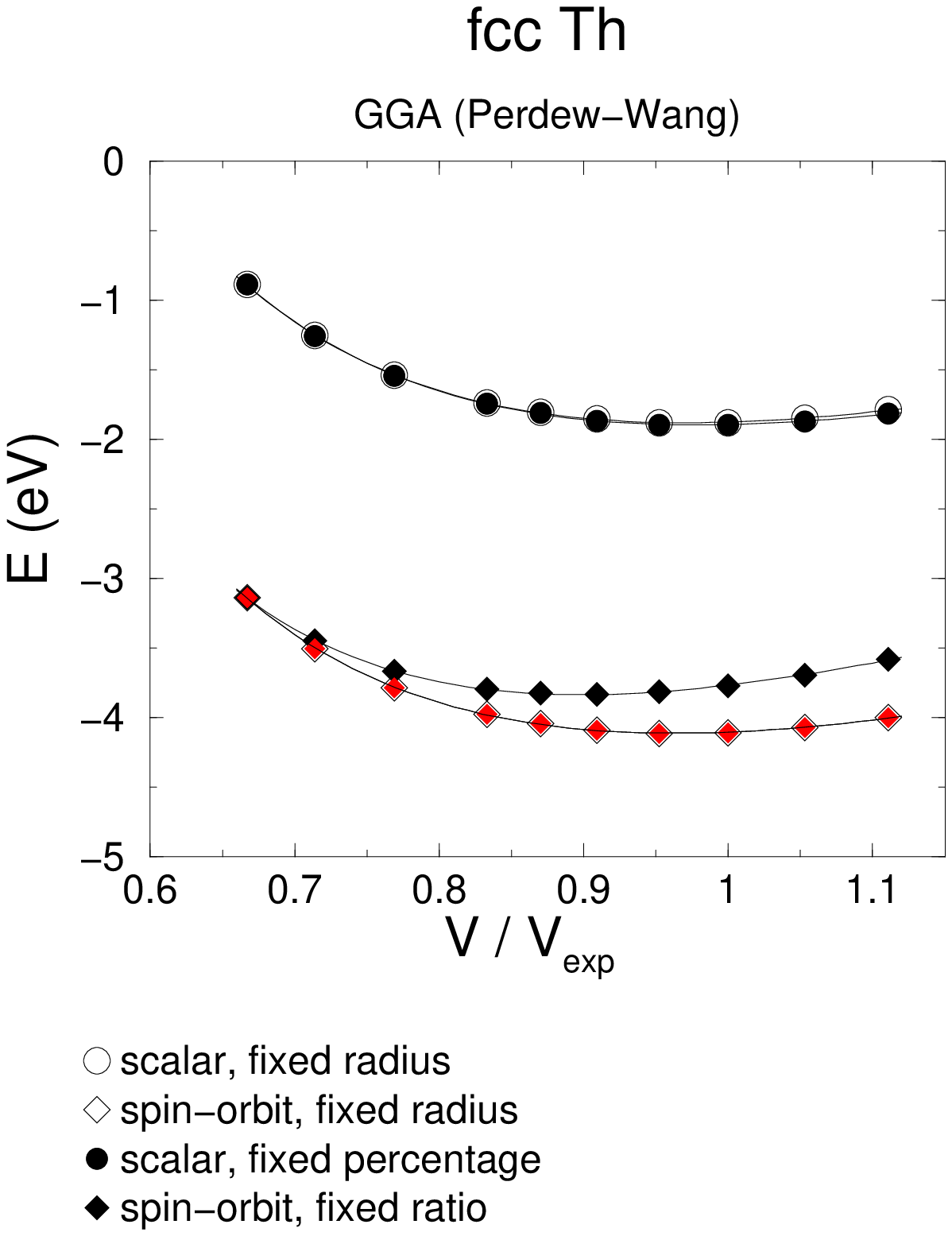,width=.60\textwidth,height=.70\textwidth}
\end{center}
\caption{
Calculated FP-LMTO total energy of Th for fixed 
muffin-tin volume and for fixed fraction of the muffin-tin volume, 
in the scalar and fully relativistic approximation.
}
\label{fig:muffin-tin}
\end{figure}

In Fig. \ref{fig:muffin-tin} we show scalar relativistic and
fully relativistic FP-LMTO total-energy calculations as a function of volume for two
types of choices of muffin-tin radius; one using a constant muffin-tin 
volume for all calculated volume/energy points and one using a constant ratio between the muffin-tin volume and unit cell volume, for all calculated points. 
On way to characterize this is to introduce the ratio between the muffin-tin radius and the Wigner-Seitz radius,
which is called  $\alpha$ in the remaining part of the manuscript. 
Hence in one of the calculations we kept $\alpha$ constant, in the manner
done
in our published FP-LMTO calculations, and in the other we varied $\alpha$ with the volume
such that the muffin-tin radius, $r_{mt}$, was kept constant. This latter approach, 
we speculate, is the approach taken by Jones {\it et al.} in their calculations,
since it is more commonly used in the LAPW method.
Note from the figure that the scalar
relativistic calculations show negligible dependence of the choice of $\alpha$
and the calculated curves lie nearly on top of each other,
whereas for the fully relativistic
calculations there is a rather large difference between the total-energy
curves, resulting in differences in equilibrium volumes of some 1-2 {\AA}$^3$.
We illustrate this further in Table \ref{tab1} were we show results from calculations
using a fixed muffin-tin radius. 
We note first that if the spin-orbit coupling is omitted 
the calculated total energy, as
well
as equilibrium volume, is the same irrespective of the choice of $r_{mt}$.
However, for the calculations including spin-orbit interaction we obtain a shift
of $\sim$ 4 \%  
for calculations including
spin-orbit interaction but with different muffin-tin radii. Hence, for certain 
choices
of muffin-tin radius (or $\alpha$) we agree with the FP-LAPW results by Jones {\it et
al.}
as we mentioned in our discussion above. Table \ref{tab1} also displays
the uncertainty
in total energy, that is up to 15 mRy/atom, which will effect the cohesive energy 
by the same
amount, depending upon choice of $r_{mt}$. 
The calculated results from the FP-LAPW method showed a similar behavior, 
i.e., when the spin-orbit coupling is considered inside the muffin-tin sphere,the calculated equilibrium volume and 
total energy depends on the choice of muffin-tin radius (see Fig.2), 
but that the calculated results are not sensitive when spin-orbit 
coupling 
is neglected. 
The numbers obtained from the FP-LAPW method are given in Table 1 and we note 
that the two methods (FP-LMTO and FP-LAPW) give very similar results.

Our conclusion from the exercise above is that the conventional wisdom still holds, FP-LMTO
and
FP-LAPW produce nearly identical results when used in a careful manner i.e.,
having converged basis functions, dense k-space summation,
using the same choice of muffin-tin radius as a function of volume etc. 
Hence, 
the conclusion of Jones {\it et al.}, that there is an error in our 
calculational procedure, is incorrect.
Nevertheless, the sensitivity of calculated volumes (to a few percent) with respect to computational parameters is unsatisfactory and it is valid 
to question, if
this signals an error in the implementation of
relativity in the current version of the FP-LMTO method, but as was discussed above,
the FP-LAPW method developed by Singh\cite{singh} and co-workers
displays similar sensitivities and
the problem seems to be
more general. Careful analysis of this problem implies that the theoretical treatment of the spin-orbit
coupling for the 6$p$ semi-core states is less accurate
in the 
implementation used in the FP-LMTO and FP-LAPW methods. 
The reason for this is that the spin-orbit coupling is larger than the 
band-width 
of the 6p band, resulting in two separated 6p$_{1/2}$ and 6p$_{3/2}$ 
bands. Since the radial basis function used for both bands is obtained from a 
differential equation with the same energy parameter, of necessity 
chosen somewhere between the 6p$_{1/2}$ and 6p$_{3/2}$ bands, a somewhat poorer treatment of these states is obtained. 
To test this hypothesis 
we have performed FP-LMTO calculations
incorporating the spin-orbit coupling for all electron states except the 6$p$
states, and the sensitivity of ground-state properties upon muffin-tin
radius disappears, confirming our hypothesis that the relativistic treatment
of the 6$p$'s does produce a dependence of the equilibrium volumes with an
uncertainty of a few percent for U, Np, and Pu and with a somewhat
 larger uncertainty
for Th. 
One could of course avoid this problem by
ignoring the spin-orbit interaction all together and obtain larger equilibrium volumes.
However, apart from not being based on theoretical grounds, such a procedure
sometimes degrades drastically the structural stability of the actinides. 
As an example we mention that the bcc-fcc energy difference of Pu (V $\sim $ 20
{\AA }$^{3}$) is 17 mRy in a scalar relativistic treatment, whereas a
calculation which includes the spin-orbit interaction gives a value of 11
mRy. In this particular case the neglect of spin-orbit 
interaction introduces an error of $\sim $
50 \%.
Although there is some sensitivity on equilibrium volumes on the choice of 
muffin-tin radius, all other calculated properties such as structural stability,
electronic structure, and transition pressures for phase transitions are
much less dependent on this choice of muffin-tin radius. 
It would of course be desirable to have a better treatment of the spin-orbit
interaction for the 6$p$ semi-core states, and to
resolve this problem one has to go one step
further and develop a full potential method using the relativistic ($j,\kappa$) basis
in which the spin-orbit coupling is implicit and exact. To date such method has
not been developed.

\section{The anomalous upturn in atomic volume for $\alpha$-Pu}

Although the main motivation for the paper by Jones {\it et al.} seems to be to
discredit previous FP-LMTO calculations they do discuss some physics
regarding the observed anomalous upturn in equilibrium volumes going from
$\alpha$-Np to $\alpha$-Pu. 
Their calculations do not show this upturn, instead 
the Np and Pu equilibrium volumes were found to be
 identical at room temperature 
(when thermal expansion was taken into account). Our previous studies of Np and
Pu show an upturn, although smaller that the observed 4\% increase
going from Np to Pu.  In Fig. \ref{fig:volumes} we show the equilibrium volumes
calculated using the FP-LAPW\cite{bs} and previous FP-LMTO calculations.
In this figure we correct for thermal expansion both for Np and Pu in the
theoretical data in order to more accurately compare with the room temperature
experimental data. This correction is especially important for Pu because this
metal has an anomalously large thermal expansion. We note that the FP-LAPW calculations
almost reproduce the experimentally observed upturn when going from Np to Pu,
and they certainly have a trend that is improved over that given by the Friedel model. The FP-LMTO
calculations reproduce the experimental trend better
although the upturn is somewhat
smaller than the experimental observation. 
It is clear from both these 
calculations that a parabolic decrease, as expected from the Friedel model of bonding
5$f$ electrons, is absent between Np and Pu. This behavior is also supported by very
recent full charge density (FCD) calculations of Vitos
{\it et al}.\cite{skriver2} that show an upturn in equilibrium volumes between Np and
Pu.
The conclusion is that it is very important to consider the correct crystal
structure of Pu and that when this is done the correct trend in equilibrium
volume is obtained.
As a whole, theoretical
calculations
(FCD and FP-LMTO) strongly suggests that the experimental upturn in the
volumes can be accounted for when careful calculations, in the correct ground-state
crystal structures and corrected for thermal expansion, are performed.
Although an improved energy functional may improve on the final few percent of the volumes of $\alpha$-Pu,
there seems to be little need in arguing for semi localized states in 
$\alpha$-Pu.\cite{bs} 

Jones {\it et al.}, argue that $\alpha$-Pu has semi localized electrons and take support in the fact that
Penicaud \cite{penicaud}, who solved the Dirac equation ($j,\kappa$)
adopting the atomic
sphere approximation (ASA), did not obtain the anomalous upturn for Pu when considering the fcc crystal structure. As has been observed many times before the
correct crystal structure must be considered to reproduce the upturn in volume between Np and Pu.
In addition Penicaud did not observe an increased volume when considering 
relativity, and the 
explanation for this result is straight-forward. Again it is 
related to the treatment of the 6$p$ semi-core states in the calculation.
Penicaud treated the 6$p$ states in a separate energy panel and for 
that reason those states were unable to
hybridize with the valence states and the hybridization mechanism
suggested by us\cite{ak3}, that explains the lowering of equilibrium volume,
is naturally absent. (We note here that the mechanism for reduction in
chemical bonding of the 5f states due to spin-orbit splitting has never been
questioned by us, we simply state that the 6$p$ contribution has an opposite
effect.) To investigate this issue further we performed calculations similar
to those of Penicaud (solving the Dirac equation and adopting the 
ASA\cite{klmto}), but with the flexibility to allow the 6$p$ states to
hybridize with the valence states. The results from these calculations
proves our argument, i.e.,
if the 6$p$'s are not allowed to hybridize with the
valence states, spin-orbit coupling does not reduce the volume
whereas if we allow for this  hybridization the reduction in 
bond distance occur.
In addition we performed calculations where the 6p states were allowed 
to hybridize, and we switched of the relativistic spin-orbit coupling of the 6p
states only. This procedure increased the equilibrium volume, in agreement with
the discussion above.

\section{conclusion}

After careful examination of the approximations involved in the FP-LAPW
and the FP-LMTO calculations,
especially regarding the 6$p$ states and the spin-orbit interaction,
we conclude that there are no errors in our previously
published FP-LMTO results for the equilibrium volume of Th or any
other element, and that the FP-LMTO and FP-LAPW method give very similar 
results when used consistently.
We have shown that both the FP-LMTO and FP-LAPW methods show some sensitivity
in calculated equilibrium volumes and total energy with respect to muffin-tin radius, which
also must be the case in the
calculations of 
Jones {\it et al.}\cite{bs}, although no such tests were reported.
Hence, the 
closer agreement with experiment in the data of Jones {\it et al.}\cite{bs}
compared to
previously published
FP-LMTO calculations, is very likely a result of 
choosing a muffin-tin radius that fortuitously gives good agreement between 
calculated and experimental volumes.
(the total energy of the two methods was unfortunately never compared in Ref.1).
This brings up an interesting point, if there is one choice of muffin-tin
radius that is better than any other. From a pragmatic point of 
view one may argue that the choice which results in best agreement with 
experiment is better, but it would be much more desirable to have a theoretical
guidance in this choice.
The rational behind the choice of a constant ratio between muffin-tin to unit cell volume 
is that one then maximizes the region in space where spin-orbit coupling is incorporated, 
not only for the lowest volume but for all volumes. 
If one 
adopts this approach, one normally considers the crystal structure with smallest
interatomic distance and makes the muffin-tin radius near touching for this geometry. 
This was done in all our calculations of equilibrium volumes as well as for 
the many successful studies of structural phase stability and pressure induced 
phase transitions in the actinides\cite{ak6}.
If one were to chose a constant muffin-tin radius and keep this radius 
the same for all volumes one could end up with a muffin-tin radius dictated by for instance the $\alpha$-Pu structure (which has a very small nearest neighbor distance) at a volume of 0.2 V$_0$ (if a large part of the equation of state
needs to be investigated), resulting in an almost vanishingly small muffin-tin
radius that when used at volumes close to the equilibrium volume ignores relativity for the majority of the unit cell. The latter approach must be less 
attractive than the former, especially for equation of state studies or investigations
of pressure induced phase transitions.
We also note that since any calculational method which treats spin-orbit coupling in an $\ell s$-basis and that includes the 6p states of the actinides as 
bands states, should 
have some dependence (a few percent) of the equilibrium volume on the choice of muffin-tin radius,
it seems at best puzzling that Jones {\it et al.} claim an accuracy 
of two different methods (one that does not construct at muffin-tin) within 1\%.
It would have been very useful to compare the total energy of the two methods
but this was unfortunately never done.

We also note here that the usefulness in LDA or GGA calculations is of course 
not the reproduction of experimental data to within the second or third
decimal point, but rather the physics that is learned from the calculation, and 
from that point of view the choice of muffin-tin radius becomes less critical 
since two different types of choices give similar trends in the equilibrium 
volume (Fig.3). However, one must be careful with these types of
calculations since due to the choice of
muffin-tin radius Jones et al. draw the erroneous conclusion that the
5f electrons in  $\alpha$-Pu are semi-localized.
There is no theoretical evidence that
there are quasi localized 5$f$-electron states in $\alpha$-Pu that causes the
experimental upturn between Np and Pu. This upturn can be reproduced from
accurate, and carefully performed, density-functional calculations,
such as the  FP-LMTO\cite{soderlind97} and FCD\cite{skriver2} methods.
Moreover, if the conclusion by Jones {\it et al.} was correct, 
the electronic structure obtained from
completely itinerant calculations would be wrong for $\alpha$-Pu and it would be 
most unlikely that such
calculations could reproduce the correct (and very complex) monoclinic ground-state
crystal structure of Pu (including pressure induced phase transitions). 
However, this has already been done in calculations by
S\"oderlind {\it et al.}\cite{soderlind97}, again suggesting that Jones {\it et al.}
have come to the wrong conclusion regarding the electronic structure of
$\alpha$-Pu. 

\section{acknowledgment}

The support from the Swedish Natural Science Foundations acknowledged. Work
performed under the auspices of the DOE.
This work
was performed under the auspices of the U.S. Department of Energy by the
Lawrence Livermore National Laboratory under contract number W-7405-ENG-48.
This work
was performed under the auspices of the U.S. Department of Energy by the
Los Alamos National Laboratory. The critical reading of the manuscript
and useful comments by Drs. H.Roeder, D.Wallace and G.Straub are greatfully acknowledged.

\newpage

\begin{table}
\begin{center}
\caption{
Equilibrium volumes and energies calculated with LDA
(Hedin-Lundqvist) as function of constant muffin-tin radius ($r_{mt}$). 
Both with and without spin-orbit interaction.
The muffin-tin radius is given in units of a$_0$ and the
volumes are given in units of a$_{0}^{3}$ and
normalized with respect to V$_{0} = 221.759$ a$_{0}^{3}$. 
Total energies are given in units of mRy and referenced 
with respect to E$_{0} = -53045146.629$ mRy.
}
\begin{tabular}{ccccc}
&\multicolumn{2}{c}{spin-orbit}&\multicolumn{2}{c}{no spin-orbit}\\
\cline{2-3}
\cline{4-5}
$r_{mt}$ & V/V$_{0}$ & E-E$_{0}$ &  V/V$_{0}$ & E-E$_{0}$ \\
      & &FP-LMTO & & \\
2.994 & 0.894 & -348.8 & 0.908 & -199.6 \\
3.085 & 0.905 & -343.2 & 0.909 & -199.9 \\
3.179 & 0.924 & -336.8 & 0.910 & -200.1 \\
3.227 & 0.935 & -333.0 & 0.912 & -200.2 \\
      & &FP-LAPW & & \\
2.46  & 0.895 & & 0.907 & \\
3.18  & 0.910 & & 0.906 & \\
\end{tabular}
\label{tab1}
\end{center}
\end{table}

\begin{figure}
\begin{center}
\epsfig{clip=,file=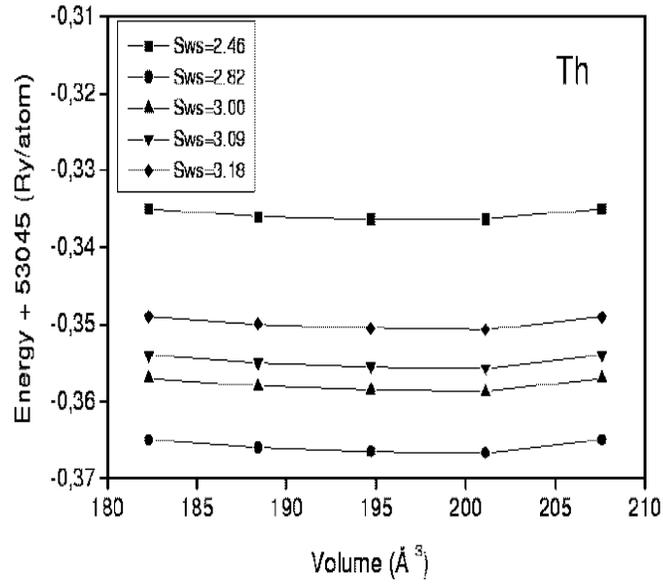,width=.55\textwidth,height=.55\textwidth}
\end{center}
\caption{
Calculated FP-LAPW total energy of Th for different but fixed 
muffin-tin volumes, 
in the fully relativistic approximation.
}
\label{fig:muffin-tin2}
\end{figure}

\begin{figure}
\label{fig:volumes}
\begin{center}
\epsfig{clip=,file=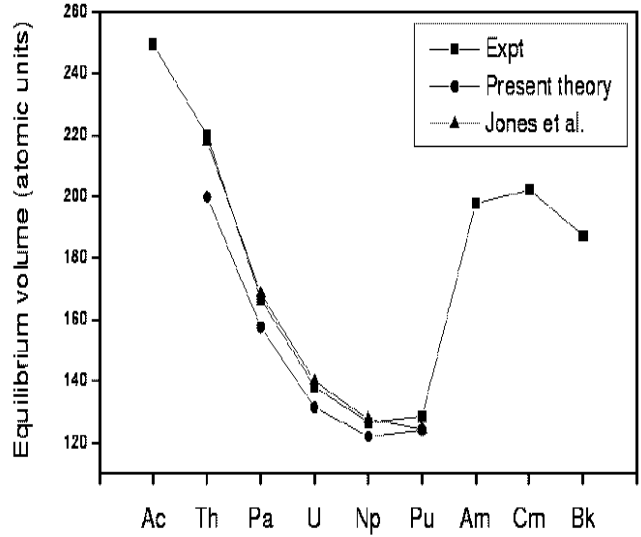,width=.55\textwidth,height=.55\textwidth}
\end{center}
\caption{%
Experimental and calculated (FP-LMTO and FP-LAPW) equilibrium 
volumes of the light actinides. The FP-LMTO calculation is by the 
authors and the FP-LAPW is calculation is from Ref.\protect\cite{bs}
} 
\end{figure}

\end{document}